\documentclass[twocolumn,pre,showpacs]{revtex4}
\usepackage{graphicx}\large
\usepackage{dcolumn}
\usepackage{amsmath}
\begin{document}

\title{Sequence-Dependent Effects on the Properties of Semiflexible Biopolymers}
\author{Zicong Zhou}
\email{zzhou@mail.tku.edu.tw} \affiliation{Department of Physics
and Graduate Institute of Life Sciences, Tamkang University, 151
Ying-chuan, Tamsui 25137, Taiwan, Republic of China} \today

\author{B\'{e}la Jo\'{o}s}
\email{bjoos@uottawa.ca}
\homepage{http://www.science.uottawa.ca/~bjoos/}
\affiliation{Ottawa Carleton Institute for Physics, University of
Ottawa Campus, Ottawa, Ontario, Canada, K1N-6N5}

\begin{abstract}
Using path integral technique, we show exactly that for a
semiflexible biopolymer in constant extension ensemble, no matter
how long the polymer and how large the external force, the effects
of short range correlations in the sequence-dependent spontaneous
curvatures and torsions can be incorporated into a model with
well-defined mean spontaneous curvature and torsion as well as a
renormalized persistence length. Moreover, for a long biopolymer
with large mean persistence length, the sequence-dependent
persistence lengths can be replaced by their mean. However, for a
short biopolymer or for a biopolymer with small persistence
lengths, inhomogeneity in persistence lengths tends to make
physical observables very sensitive to details and therefore less
predictable.
\end{abstract}
 \maketitle

\section{Introduction}
The conformal and mechanical properties of double-stranded DNA
(dsDNA) have attracted considerable attention due to the central
role that dsDNA plays in biological processes. Recent progresses
in experimental techniques such as laser or magnetic tweezers,
atomic force microscopy, and other single molecule techniques make
it possible to manipulate and observe single biomolecules directly
\cite{SFB92,SABBC96,CLHLVCC96,SCB96}, allowing better comparisons
between theoretical predictions and experimental observations. In
theoretical studies, a semiflexible biopolymer is often modelled
as a filament. The simplest model for a filament, called the
wormlike chain (WLC) model, views the filament as an inextensible
continuous chain with a uniform bending rigidity but with
vanishing cross section, and has been successfully applied to the
entropic elasticity of dsDNA \cite{KP49,MS94,BMSS94,MS95}.
Furthermore, the wormlike rod chain (WLRC) model which regards the
filament as a chain with spontaneous twist and a finite circular
cross section, has been used to explain the supercoiling property
of dsDNA \cite{MS94,BMSS94,MS95,FRO97}. Owing to the importance of
DNA, recently there has been a lot of theoretical work on the WLC
and WLRC models as well as their modifications and extensions
\cite{KP49,MS94,BMSS94,MS95,FRO97,GT97,KR03,PR00,PR01,
ZLJ05,ZJLYJ07,WN07,PN98,BDM98,RR04,ZSCGSS01,NHJWKD03,PT07,MFFA07,ZZ07}.

Traditional models of filaments are essentially homogeneous. In
other words, these models are defined by $s$-independent
parameters, where $s$ is the arclength. However, biopolymers are
often sequence-dependent and so are heterogeneous. Several recent
works have revealed that the sequence-disorder has remarkable
effects on the properties of dsDNA
\cite{PN98,BDM98,RR04,ZSCGSS01,NHJWKD03,PT07,MFFA07}. Based on the
elastic model, two effects of sequence-disorder have to be
considered. First, structural inhomogeneity results in variations
of the bending rigidity along the chain, and can be described by
the $s$-dependent persistence length $l_p(s)$
\cite{PT07,NHJWKD03}. It has been demonstrated that for a long DNA
chain without long-range correlation (LRC) in $l_p(s)$, this
effect can be well accounted for by a simple replacement of the
uniform persistence length $l_p$ in the WLC model by a proper
average of the $l_p(s)$ \cite{PT07}. However, for loop formation
in a short DNA chain this effect becomes complex because the
looping probability of a typical filament segment is not a
well-defined function of its length \cite{PT07}. Secondly, the
local structure of the dsDNA can be characterized by the
sequence-dependent spontaneous curvature $\kappa_0(s)$
\cite{PN98,BDM98,RR04,PT07,MFFA07,ZSCGSS01,NHJWKD03}. For short
dsDNA chains, special sequence order may favor a macroscopic
spontaneous curvature \cite{SFCTFMWW06,HLDH97,DPGHH96}. On the
other hand, for long dsDNA chains, the effects of $\kappa_0(s)$ is
dependent on the degree of correlation in basepairs. Without
correlation or with short range correlation (SRC), the effect can
be also reduced into a renormalization of $l_p$ in the WLC model
\cite{PT07,PN98,BDM98,MFFA07}. However, with LRC, the simple
correction to the $l_p$ is invalid because the biopolymer develops
a macroscopic intrinsic curvature \cite{MFFA07}. Moreover,
computer simulations suggest that the mean of $\kappa_0(s)$,
rather than the details of its distribution, determines the
looping probability of a filament \cite{RR04}. However, all
analytical approaches on the sequence-dependent effects are
limited to specified properties and on a WLC-based model with
vanishing intrinsic curvature and with weak or vanishing external
force, a rigorous proof on the general elastic continuous model is
yet elusive. Bearing in mind that many dsDNA possess macroscopic
intrinsic curvature \cite{SFCTFMWW06,HLDH97,DPGHH96,MFFA07}, an
analytical approaches on the general model is of special
important.

\section{Model}
Using $s$ as variable, the configuration of a filament can be
described by a triad of unit vectors $\{{\bf t}_i(s)\}_{i=1,2,3}$,
where ${\bf t}_3\equiv d{\bf r}(s)/ds$ is the tangent to the
center line {\bf r}($s$) of the filament, and {\bf t}$_1$ and {\bf
t}$_2$ are oriented along the principal axes of the cross section.
The orientation of the triad as one moves along the filament is
given by the solution of the generalized Frenet equations that
describe the rotation of the triad vectors \cite{PR00,PR01,KR03},
$d{\bf t}_i(s)/ds=-\Sigma_{j,k}\epsilon_{ijk} \omega_j(s)${\bf
t}$_k(s)$, where $\epsilon_{ijk}$ is the antisymmetric tensor, and
\{$\omega_i(s)$\} are the curvature and torsion parameters.

The elastic energy of a filament with $s$-dependent spontaneous
curvatures $\zeta_{1}(s)$, $\zeta_{2}(s)$, spontaneous twist rates
$\zeta_{3}(s)$ and persistence lengths $a_i(s)$ can be written as
\cite{PR00,PR01,KR03}
\begin{eqnarray}
{E \over k_B T}\equiv{\cal E}={1\over 2}\int_0^L \sum_{i=1}^3
a_i(\omega_i-\zeta_{i})^2ds,\label{energy3D}
\end{eqnarray}
where $T$ is the temperature, $k_B$ is the Boltzmann constant, and
$L$ is the total arclength of the filament and is a constant so
that the filament is inextensible.

If $\zeta_i$ and $a_i$ are well-defined (i.e., without randomness)
functions of $s$, a macroscopic quantity $B$ is defined as the
average with Boltzmann weights over all possible conformations, so
is a path integral in the form \cite{PR00,PR01,KR03}
\begin{eqnarray}
B &\equiv &\left< B[\{\omega_i(s)\}]\right>= { \int {\cal
D}[\omega_i] B[\{\omega_i(s)\}] \text{e}^{-\cal E}\over \int {\cal
D}[\omega_i] \text{e}^{-\cal E}}. \label{mean1}
\end{eqnarray}
Function $B[\{\omega_i(s)\}]$ represents different physical
situations. For instance, if $B[\{\omega_i(s)\}]={\bf
t}_j(s_1)\cdot {\bf t}_k(s_2)$, we find the orientational
correlation function between ${\bf t}_j$ and ${\bf t}_k$; if
$B[\{\omega_i(s)\}]=|{\bf r}_L-{\bf r}_0|^2$, we obtain the
end-to-end distance, where ${\bf r}_L={\bf r}(L)$ and ${\bf
r}_0={\bf r}(0)$; if $B[\{\omega_i(s)\}]=\delta({\bf
r}-\int_0^L{\bf t}_3ds)$, we get the distribution function of
end-to-end vector. The applied force can be evaluated using this
distribution function; if $B[\{\omega_i(s)\}]=\delta({\bf
r}_L-{\bf r}_0)\delta[{\bf t}_{3}(L)-{\bf t}_{3}(0)]$, we find the
looping probability. Note that $B[\{\omega_i(s)\}]$ may be a very
complex function of $\omega_i(s)$, but its detailed form is
irrelevant in this work since it is independent on $a_i$ and
$\zeta_i$. If both ends are free of external force, $B$ represents
the intrinsic property of the system. On the other hand, if we fix
both ends of the filament, we obtain quantity in the constant
extension ensemble.

\section{The effects of the sequence-dependent spontaneous
curvatures and torsions}
We first consider the effects of the
$\zeta_i(s)$ alone but leave $a_i$'s as well-defined. For a
biopolymer without correlation on $\zeta_i(s)$, or with SRC but in
the coarse-grained model, the distribution of $\zeta_i(s)$,
$W(\{\zeta_{i}\})$, can be written as a Gaussian distribution with
nonvanishing average $\omega_{i0}$
\begin{eqnarray}
W(\{\zeta_{i}\})= \text{exp}\left[ -\sum_{i=1}^3\int {k_i\over 2
}(\zeta_i(s)-\omega_{i0})^2 ds\right]. \label{weight1}
\end{eqnarray}
In other words, $\zeta_i(s)$'s are delta correlated along the
chain:
\begin{eqnarray}
\left< [\zeta_i(s)-\omega_{i0}]
[\zeta_j(s')-\omega_{j0}]\right>={1\over k_i} \delta_{ij}
\delta(s-s').
\end{eqnarray}
In this case, we need to average over $\zeta_i$ again for $B$ so
\begin{eqnarray}
B_\zeta={\int{\cal D}[\zeta_i]W(\{\zeta_{i}\})\left({\int {\cal
D}[\omega_i]\text{e}^{-\cal E} B[\{\omega_i(s)\}]\over {\cal
Z}_\omega }\right) \over {\cal Z}_\zeta}, \label{mean2}
\end{eqnarray}
where ${\cal Z}_\omega \equiv \int {\cal
D}[\omega_i]\text{e}^{-\cal E}$ and ${\cal Z}_\zeta \equiv
\int{\cal D}[\zeta_i]W(\{\zeta_{i}\})$ are essentially Gaussian
integrals so are independent of $\zeta_i$ or $\omega_{i}$ but
dependent on $a_i$ or $k_i$, respectively. Now using the identity
\begin{eqnarray}
&&{ \int{\cal D}[\zeta] \text{e}^{-{1\over 2}\int
[a(\omega-\zeta)^2+k (\zeta-\omega_0)^2]ds} \over \int{\cal
D}[\zeta]\text{e}^{-{1\over 2}\int k (\zeta-\omega_0)^2ds}
}\nonumber \\&=&\text{e}^{-\int {\beta \over
2}[\omega(s)-\omega_0]^2ds}{ \int{\cal D}[\omega] \text{e}^{-\int
{a\over 2}(\omega-\zeta)^2ds} \over \int{\cal D}[\omega
]\text{e}^{-\int {\beta \over 2}(\omega-\omega_0)^2ds} },
\label{identity1}
\end{eqnarray}
and exchanging the order in integral, we finally obtain
\begin{eqnarray}
B_\zeta &=& {\int{\cal D}[\zeta_i]W(\{\zeta_{i}\})\left( \int
{\cal D}[\omega_i]\text{e}^{-\cal E} B[\{\omega_i(s)\}]\right)
\over
{\cal Z}_\omega \text{ }{\cal Z}_\zeta}\nonumber \\
&=&{1\over {\cal Z}_\omega}\int {\cal D}[\omega_i]
B[\{\omega_i(s)\}]{\int{\cal D}[\zeta_i] \text{e}^{-\cal E}
W(\{\zeta_{i}\}) \over {\cal Z}_\zeta}\nonumber \\
&=& { \int {\cal D}[\omega_i] B[\{\omega_i(s)\}] \text{e}^{-{\cal
H}}\over \int {\cal D}[\omega_i] \text{e}^{-{\cal H}}},
\label{mean3}
\end{eqnarray}
where
\begin{eqnarray}
{\cal H}={1\over 2}\int_0^L \sum_{i=1}^3
\beta_i[\omega_i(s)-\omega_{i0}]^2ds,\label{energy3}
\end{eqnarray}
and $\beta_i=a_i k_i/(a_i + k_i)$. Note that Eq. (\ref{mean3}) is
valid for any length, and even if $a_i$, $k_i$ and $\omega_{i0}$
are $s$-dependent. Comparing Eqs. (\ref{mean1}) and (\ref{mean3}),
we reach the conclusion that the effects of $\zeta_i(s)$ can be
incorporated into a model with well-defined mean spontaneous
curvatures and torsion $\omega_{i0}$ as well as renormalized
persistence lengths $\beta_i$. This conclusion agrees with what
has been found in the special case with $k_1=k_2=l_p^s$ and
$\omega_{10}=\omega_{20}=0$, namely that the randomness of the $\zeta_i(s)$
can be accounted for by replacing $l_p$ with an effective
persistence length $l_p^{\text{eff}}$ in the WLC model, where
$1/l_p^{\text{eff}}=1/ l_p+1/l_p^s$ \cite{PN98,PT07}. A different
form is obtained with a half-Gaussian distribution of disorder on
curvature, which yields $l_p^{\text{eff}}=l_p\left(1-{1\over
2}\sqrt{l_p/l_p^s}\right)$ \cite{BDM98}. Our result also agrees
with the conclusion obtained from computer simulation that the
mean spontaneous curvature, rather than the details of its
distribution, determines the looping probability of a filament
\cite{RR04}. We should note that the proofs in Refs.
\cite{PN98,BDM98,PT07} are limited to the special case with
$a_1=a_2$, $\omega_{10}=\omega_{20}=0$, and under weak or
vanishing external force, but our proof is rigorous and generally
valid.

The next question is would it be possible to replace the
nonvanishing $\omega_{i0}$ in the model by $\omega_{i0}=0$ by
renormalizing further $\beta_i$? The answer to this question
depends on the situation. For the end-to-end distance of a very
long filament free of external force, the answer is yes
\cite{PR00,PR01,ZZ07}. But the convergence to that limit is slow
so the above replacement is poor for moderate length (from a few
$l_p$ to about 20 $l_p$) two-dimensional filaments \cite{ZZ07}.
When relating applied force and extension, such a replacement is
also only reasonable at low force and large $L$ \cite{MS95,ZZ07}.

\section{The effects of the sequence-dependent persistence lengths}
Now we consider the effects of the $a_i(s)$ alone but keep
$\zeta_i$ as well-defined. In this case, we assume that the
distribution of the $a_i$ is half-Gaussian since $a_i<0$ is
meaningless:
\begin{eqnarray}
\rho(\{a_i(s)\})=\text{exp}\left[ -\sum_{i=1}^3\int {b_i\over 2 }
[a_i(s)-\bar{a}_i]^2 ds\right], \text{ }a_i>0. \label{weight3}
\end{eqnarray}
It is difficult to do an average over $a_i$ if $\bar{a}_i$ is
small. Therefore, we assume that $\bar{a}_i$ is far from zero,
which is reasonable for semiflexible biopolymers such as dsDNA, so
approximately we have
\begin{eqnarray}
{\cal Z}_a&\equiv& \int_0^\infty  {\cal
D}[a_i]\rho(\{a_i(s)\})\nonumber \\ &\approx& \int_{-\infty
}^\infty {\cal D}[a_i]\text{exp}\left[ -\sum_{i=1}^3\int {b_i\over
2 }a_i^2(s) ds\right],
\end{eqnarray}
and so ${\cal Z}_a$ is dependent on $b_i$ only. In this case,
\begin{eqnarray}
B_a&=&{1\over {\cal Z}_a}\int{\cal D}[a_i]\rho(\{a_i(s)\})\left[{
\int {\cal D}[\omega_i] B[\{\omega_i(s)\}] \text{e}^{-{\cal
E}}\over {\cal Z}_\omega'}\right], \nonumber \\
&=& {1\over {\cal Z}_a}\int{\cal D}[\omega_i] B[\{\omega_i(s)\}]
{\cal C}[\{\omega_i(s) \}],\label{mean5}
\end{eqnarray}
where
\begin{eqnarray}
{\cal C}[\{\omega_i(s) \}]= \int {\cal D}[a_i]{\rho(\{a_i(s)\})
\text{e}^{-{\cal E}}\over {\cal Z}_\omega' },\label{mean6}
\end{eqnarray}
and ${\cal Z}_\omega'=\int {\cal D}[\omega_i]\text{e}^{-{\cal E}}$
is dependent on $a_i(s)$. Applying standard path integral methods
\cite{KLB93} leads to
\begin{eqnarray}
{\cal Z}_\omega'\propto \lim_{N \rightarrow \infty }
\prod_{i=1,3}\prod_{j=1,N}\left( {\pi \over 2 a_{ij} \epsilon }
\right)^{N/2},\label{part6}
\end{eqnarray}
where $\epsilon=L/N$, and $a_{ij}=a_i[(j-1)\epsilon ]$ is the
discretized $a_i(s)$. The form of ${\cal Z}_\omega'$ makes it
impossible to find a closed form for ${\cal C}[\{\omega_i(s) \}]$.
However, if the distribution in $a_i$ is narrow, which should be
the case when the molecules forming the different segments are
similar such as dsDNA, we can then replace the $a_i$ in Eq.
(\ref{part6}) by $\bar{a}_i$, so ${\cal Z}_\omega'$ can be taken
out of the integrand in ${\cal C}$ [see Eq. (\ref{mean6})] and
written as
\begin{eqnarray}
{\cal Z}_\omega'&\approx& \int {\cal D}[\omega_i]\text{e}^{{-\cal
E}_1}, \\ \mbox{where} \ \ \ {\cal E}_1&=&{1\over 2}\sum_{i=1}^3
\int_0^L \left[\bar{a}_i(\omega_i-\zeta_{i})^2\right]ds
\label{energy3D1}.
\end{eqnarray}
As a consequence,
\begin{eqnarray}
{\cal C}[\{\omega_i(s) \}]\approx  {1 \over {\cal Z}_\omega' }\int
\cal D}[a_i]{\rho(\{a_i(s)\})\text{e}^{-{\cal E}}.\label{mean7}
\end{eqnarray}
Now using the identity
\begin{eqnarray}
&&b\left[a-\bar{a}+{1\over
2b}(\omega-\zeta)^2\right]^2+\bar{a}(\omega-\zeta)^2-{1\over
4b}(\omega-\zeta)^4\nonumber
\\&=&b(a-\bar{a})^2+a(\omega-\zeta)^2,
\end{eqnarray}
we obtain
\begin{eqnarray}
{\cal C}[\{\omega_i(s) \}]&\approx &{{\cal Z}_a\over {\cal Z}_\omega'}
\text{e}^{-{\cal E}_2},\\
B_a&\approx &{ \int {\cal D}[\omega_i] B[\{\omega_i(s)\}]
\text{e}^{-{\cal E}_2}\over \int {\cal D}[\omega_i]
\text{e}^{-{\cal E}_1}}, \label{mean8}
\end{eqnarray}
where
\begin{eqnarray}
{\cal E}_2&=&{1\over 2}\sum_{i=1}^3 \int_0^L
\left[\bar{a}_i(\omega_i-\zeta_{i})^2-{1\over
4b_i}(\omega_i-\zeta_i)^4\right]ds.\label{energy3D2}
\end{eqnarray}
Due to the existence of the term $(\omega_i-\zeta_i)^4$ in Eq.
(\ref{energy3D2}), $B_a$ is divergent if there is no constraint on
$\omega_i$. However, biopolymers cannot have infinite $\omega_i$,
so there is a cutoff for $\omega_i$. This cutoff should be large
enough so that for the $(\omega_i-\zeta_{i})^2$ term we can remove
the constraint on $\omega_i$. Moreover, it was reported that for a
dsDNA chain with 64 trinucleotides,
$\left<[l_p^{-1}(s)-l_p]^2\right>\approx 0.13 l_p^{-2}$
\cite{PT07}. This means that even for a short dsDNA chain, the
distribution of $l_p$ is not very wide. It is therefore reasonable
to expect that for a long semiflexible biopolymer, the
distribution of $a_i$'s becomes very sharp and the $b_i$'s are
large. We can then expect that the $(\omega_i-\zeta_i)^4$ term
remains small and can be neglected up to the cutoff of $\omega_i$.
Consequently we have
\begin{eqnarray}
B_a&\approx &{ \int {\cal D}[\omega_i] B[\{\omega_i(s)\}]
\text{e}^{-{\cal E}_1}\over \int {\cal D}[\omega_i]
\text{e}^{-{\cal E}_1}}. \label{mean9}
\end{eqnarray}
Eq. (\ref{mean9}) means that we can replace $a_i(s)$ by
$\bar{a}_i$. This conclusion agrees with the conclusion for the
special case $a_1=a_2$, $a_3=0$ and $\omega_{i0}=0$ \cite{PT07}.
However, for a short biopolymer, the contribution from
$(\omega_i-\zeta_i)^4$ cannot be ignored, and the results tend to
be divergent making the averages poorly defined functions of $L$,
as was reported for the special case \cite{PT07}.

From the above derivations, we see that it is not a simple task to
study the combined effects of the sequence-dependence of  $a_i$
and $\zeta_i$ because of the term $(\omega_i-\zeta_i)^4$ and the
fact that $\beta_i$ is not a linear function of $a_i$. However,
when Eq. (\ref{mean9}) is valid, Eqs.
(\ref{mean3})-(\ref{energy3}) can be recovered with the
replacement of $a_i$ by $\bar{a}_i$.

\section{Conclusion and discussion}
In summary, we present a rigorous and general proof that for a
biopolymer without correlation or with SRC on spontaneous
curvatures and torsions $\zeta_i$, the effects of
sequence-disorder on $\zeta_i$ can be incorporated into a model
with well-defined mean $\zeta_i$ [i.e. $\omega_{i0}$] as well as
renormalized persistence length, no matter how long the biopolymer
and how large the external force may be. Moreover, if the
biopolymer is sufficiently long and has a large enough mean
persistence length, the sequence-dependent persistence length
$a_i(s)$ can be replaced by its mean $\bar{a}_i$. Note that
``semiflexible" in general means that the biopolymer has a
sufficiently large $\bar{a}_i$, our above conclusions can be
safely applied to long semiflexible biopolymers such as dsDNA.
However, for a short biopolymer or for a biopolymer with small
$\bar{a}_i$, the effects of inhomogeneity in $a_i(s)$ become very
complex and tend to make physical observables very sensitive to
the details of $a_i(s)$. Our derivations are quite general, so the
conclusions can be applied to various conformal and mechanical
properties.

We also should remind that our proof works only in the constant
extension ensemble. But it is reasonable to expect that these
conclusions also can be applied to sufficient long biopolymers
since in this case the structural details must be immaterial. It
has been known that the constant extension ensemble and the
constant force ensemble may be inequivalent at finite $L$. For
constant force ensemble, the same conclusion has been achieved at
a special case with $a_1=a_2$, $\omega_{10}=\omega_{20}=0$ and
under weak applied force \cite{PN98}, but a proof for the general
case is not yet available. In constant force ensemble, we need to
add a term, which is the contribution of the external force, into
the energy, and the energy becomes \cite{PR01,KR03,ZLJ05,ZJLYJ07}
\begin{eqnarray}
{\cal E}=\int_0^L \left[{1\over 2}\sum_{i=1}^3
a_i(\omega_i-\zeta_{i})^2-F \cos\theta \right]ds,\label{eforce}
\end{eqnarray}
where $F\equiv f/(k_B T)$ and $f$ is the applied force. $\theta$
is the angle between force and the tangent of the central line of
the filament and is a very complicate function of $\omega_i$. This
force term makes $Z_\omega$ dependent on $\zeta$ and so renders
the exchange of the order in integral [from Eq. (\ref{mean2}) to
Eq. (\ref{mean3})] illegal. Therefore, whether the same conclusion
is valid in the constant force ensemble for a short biopolymer is
still an open question.

Moreover, we should recall that SRC in this work means that with
proper length scale, the distribution function is Gaussian. What
is the proper length scale is not yet very clear. It has been
reported that the most bendable DNA sequences are those that wrap
around nucleosomes, and there exists a correlation in the way they
are arranged. Along the DNA contour, AA/TT/TA dinucleotides have a
periodicity of about 10 basepairs and this is the signature of the
region with high affinity to nucleosomes \cite{SFCTFMWW06}.
Therefore, a reasonable estimate of the proper length scale for
dsDNA is about 10 basepairs. We do not consider systems with LRC
in $\zeta_i$ and/or $a_i$ in this work so it deserves further
investigation. But we should point out that LRC in sequences is
not the same as LRC in $\zeta_i$. For instance, for a homopolymer,
the correlation in sequences is $100\%$, but it can be described
by constant (or vanishing) $\zeta_i$ so can be regarded as no
correlations in $\zeta_i$ since it corresponds to the limit case
of Gaussian distribution with vanishing variances. In the more
general case, LRC in sequences tends to make neighbor sequences
have similar bending so to develop a macroscopic intrinsic
curvature, and the local intrinsic curvatures may have only a
small random deviation from its mean, it in turn leads to the SRC
in $\zeta_i$, at least in the first approximation. As a
consequence, many properties, such as the behavior of the
end-to-end distance \cite{ZZ07}, of such a biopolymer can be well
accounted for by a model with constant (or well-defined)
spontaneous curvature. Finally, the method used in this work may
be applied to some other similar systems, such as Hookian springs
with random natural lengths, or a quantum harmonic oscillator with
randomly moving centers, or a quantum planar rotor in a randomly
rotating coordinate system.

\section* {Acknowledgments}
This work has been supported by the National Science Council of
the Republic of China under grant no. NSC 96-2112-M-032-002, the
Physics Division, National Center for Theoretical Sciences at
Taipei, National Taiwan University, Taiwan, ROC, and the Natural
Sciences and Engineering Research Council of Canada.

\end{document}